\documentclass[11pt,twoside]{article}  
\usepackage{apn3conf}
\usepackage{epsf}

\begin{document}   

\title{Grain Size Distributions and Photo-Electric Heating in Ionized Regions}
\titlemark{Photo-Electric Heating in Ionized Regions}

\author{P. A. M. van Hoof}
\affil{Queen's University Belfast, APS Division, Physics Dept., Belfast, BT7~1NN, Northern Ireland}
\author{J. C. Weingartner\altaffilmark{1}, P. G. Martin}
\affil{Canadian Institute for Theoretical Astrophysics, 60 St. George Street, Toronto, ON M5S~3H8, Canada}
\author{K. Volk}
\affil{University of Calgary, 2500 University Dr. NW, Calgary, AB T2N~1N4, Canada}
\author{G. J. Ferland}
\affil{University of Kentucky, 177 CP Building, Lexington, KY 40506, USA}
\altaffiltext{1}{Current address: Department of Physics and Astronomy, George Mason 
University, 4400 University Drive, MSN 3F3, Fairfax, VA  22030}

\contact{Peter van Hoof}
\email{p.van-hoof@qub.ac.uk}

\paindex{van Hoof, P. A .M.}
\aindex{Weingartner, J. C.}
\aindex{Martin, P. G.}
\aindex{Volk, K.}
\aindex{Ferland, G. J.}

\authormark{van Hoof et al.}

\setcounter{footnote}{1}


\begin{abstract}          
In this paper we present results obtained with the new grain code in Cloudy
which underline the strong effect of photo-electric heating by grains in
photo-ionized regions. We will study the effect that the distribution of grain
sizes has on the magnitude of the effect, and show that this effect is nothing
short of dramatic. This makes the grain size distribution an important
parameter in modeling of photo-ionized regions such as H\,{\sc ii} regions and
planetary nebulae.
\end{abstract}

\section{Introduction}

This paper focuses on the grain model in Cloudy, which has undergone a major
upgrade in the last couple of years. The first grain model was introduced to
Cloudy in 1990 to facilitate more accurate modeling of the Orion nebula (for a
detailed description see Baldwin et al. 1991). In subsequent years, this model
has undergone some revisions and extensions, but remained largely the same.
Recently, our knowledge of grains has been greatly advanced by the results
from the {\it ISO} mission. In view of these rapid developments we have undertaken a
major upgrade of the grain model in Cloudy. The two main aims were to make the
code more flexible and versatile, and to make the modeling results more
realistic. These are the main improvements:

\begin{itemize}
\item
We have included a Mie code for spherical particles. The necessary optical
constants needed to run the code are read from a separate file.
\item
A mixing law has been introduced to the code. This allows the user to define
grains which are mixtures of different materials.
\item
The absorption and scattering opacities can be calculated for completely
arbitrary grain size distributions (including single-sized grains).
\item
The size distribution can be resolved in many small bins (the user can choose
how many), and all physical quantities are calculated for each bin separately.
This allows non-equilibrium heating to be treated correctly for the smallest
grains in the size distribution, and more realistic grain emission spectra to
be calculated. It will also improve modeling of the photo-electric effect
which also depends strongly on grain size.
\item
The code for non-equilibrium treatment of PAH's has been extensively rewritten
for the new grain model. It now works automatically and efficiently with all
grain types and sizes, under all conditions.
\item
We have updated the grain physics following the discussion in Weingartner \&
Draine (2001). This includes an improved treatment of the photo-electric
effect and the electron sticking probability. The code now also uses discrete
charge states for the grains, but our treatment deviates somewhat from
Weingartner \& Draine (2001) in that we use the hybrid grain charge model (van
Hoof et al., 2001), instead of a fully resolved non-equilibrium charge
distribution. We have shown that the hybrid grain charge model is sufficiently
accurate for all realistic astronomical applications.
\end{itemize}

The new grain model is currently being distributed as part of the beta release
of Cloudy 96. Cloudy can be obtained from \htmladdnormallinkfoot{the Cloudy
website}{http://www.nublado.org/}. The Cloudy 96 beta release can currently
be found under ``Other versions''.

\section{Photo-electric Heating and Grain Size Distributions}

\begin{figure}
\centerline{\epsfxsize=\textwidth\epsfbox[12 347 558 721]{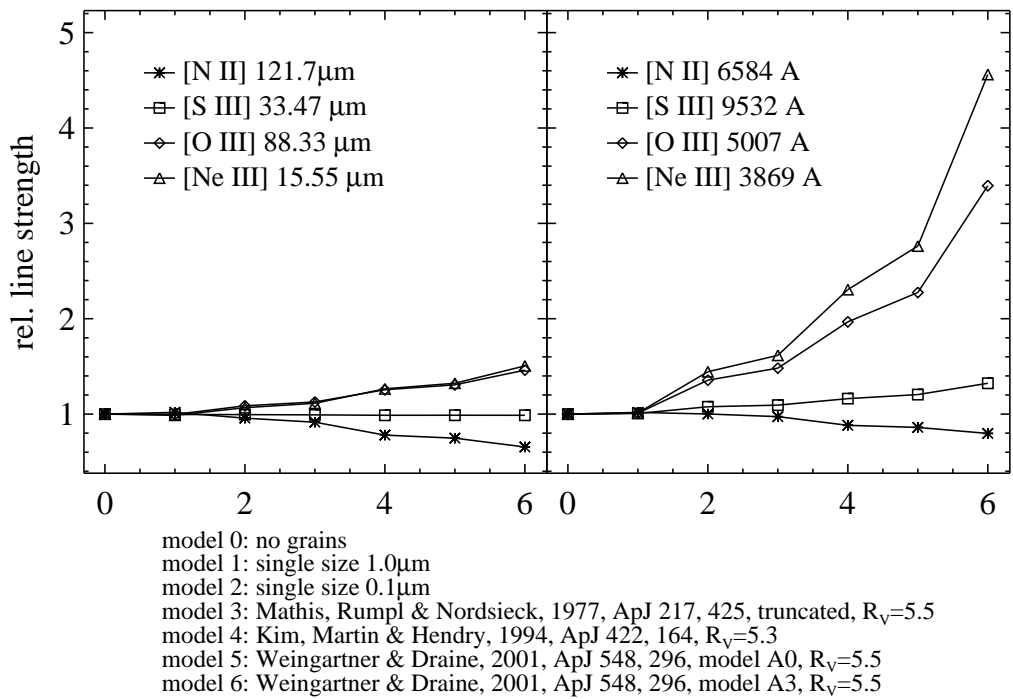}}
\vspace{6mm}
\centerline{\epsfxsize=0.7\textwidth\epsfbox[53 353 507 684]{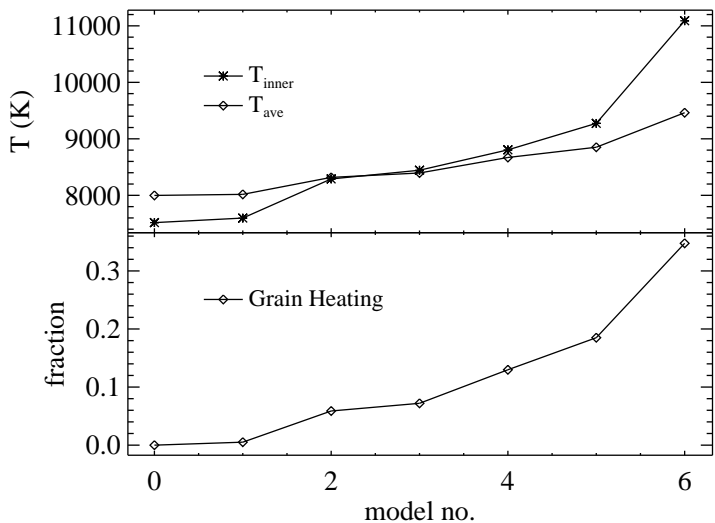}}
\caption{Paris H\,{\sc ii} region models. In the top left panel we show the
relative line strengths for selected infrared fine-structure lines. These are
expected to be mostly insensitive to electron temperature and therefore show
the difference in the overall ionization structure. The line strengths are
normalized to the line strength in the dust-free model. In the top-right panel
we show optical/UV forbidden lines of the same species. In the bottom panels
we show the electron temperature at the inner edge, as well as averaged over
the ionized region, and the fraction of the total gas heating that is due to
the photo-electric effect.}
\label{fig:hii}
\end{figure}

\begin{figure}
\centerline{\epsfxsize=\textwidth\epsfbox[12 347 558 721]{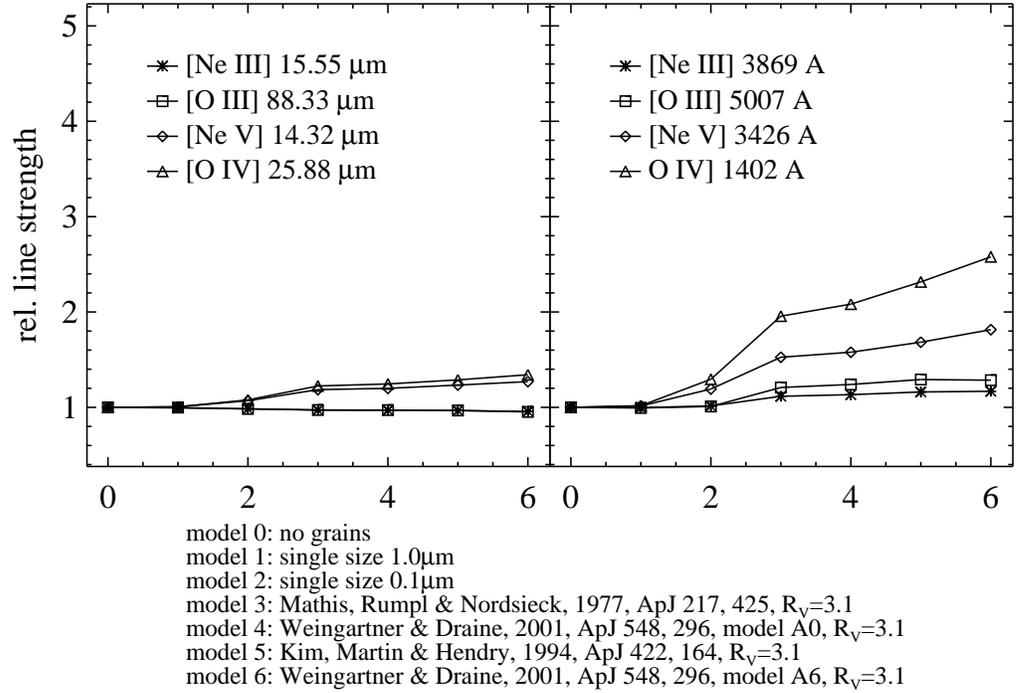}}
\vspace{6mm}
\centerline{\epsfxsize=0.7\textwidth\epsfbox[53 353 507 684]{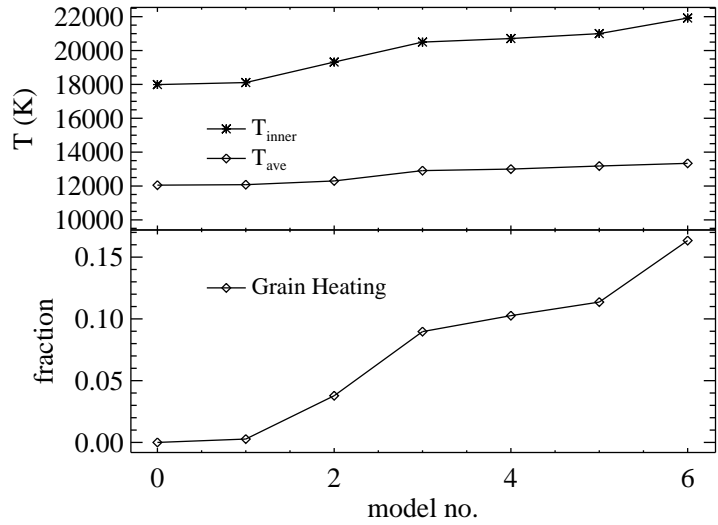}}
\caption{Same as Fig.~\ref{fig:hii}, but for the Paris PN models with graphite only.}
\label{fig:pn:gra}
\end{figure}

\begin{figure}
\centerline{\epsfxsize=\textwidth\epsfbox[12 347 558 721]{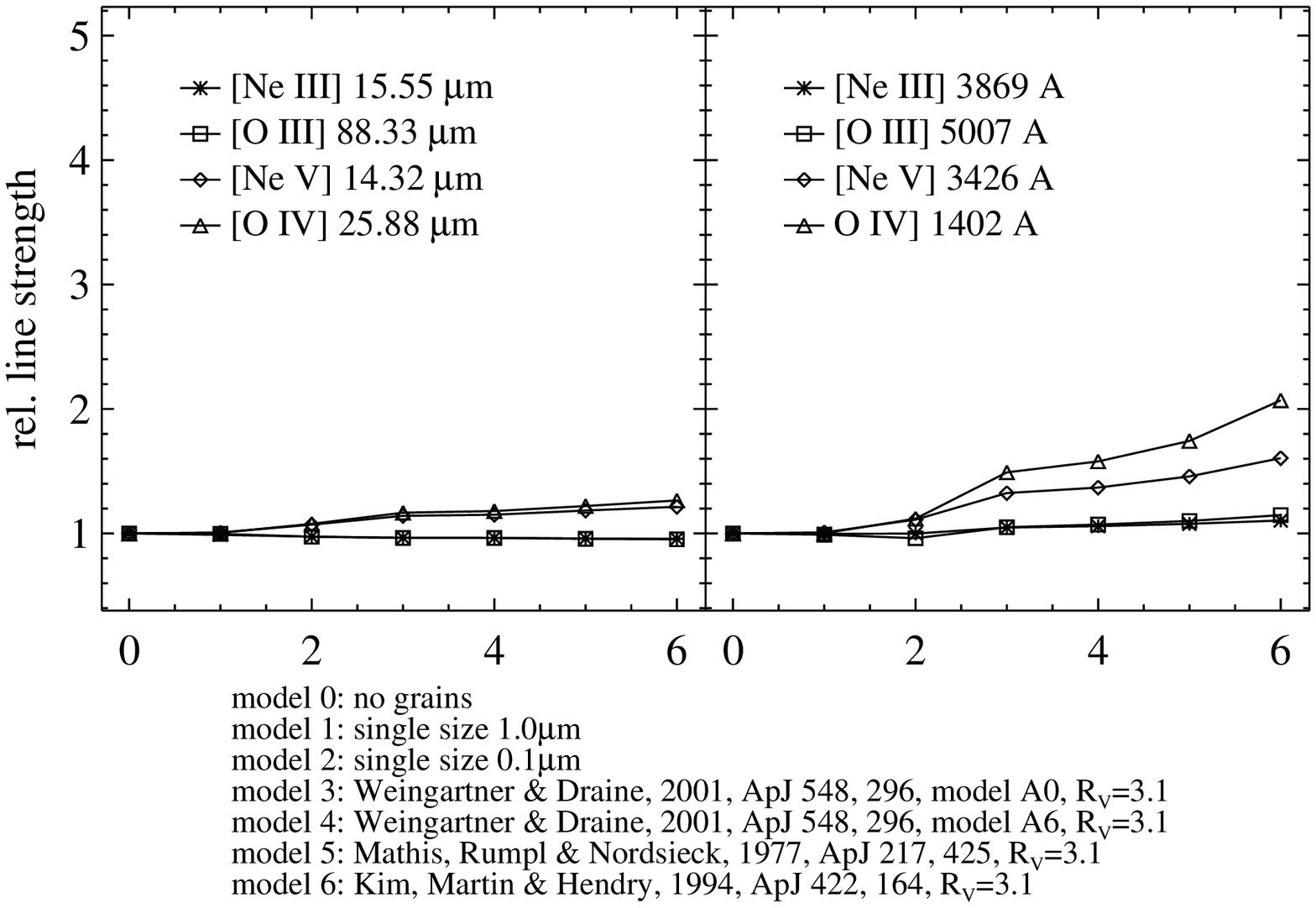}}
\vspace{6mm}
\centerline{\epsfxsize=0.7\textwidth\epsfbox[53 353 507 684]{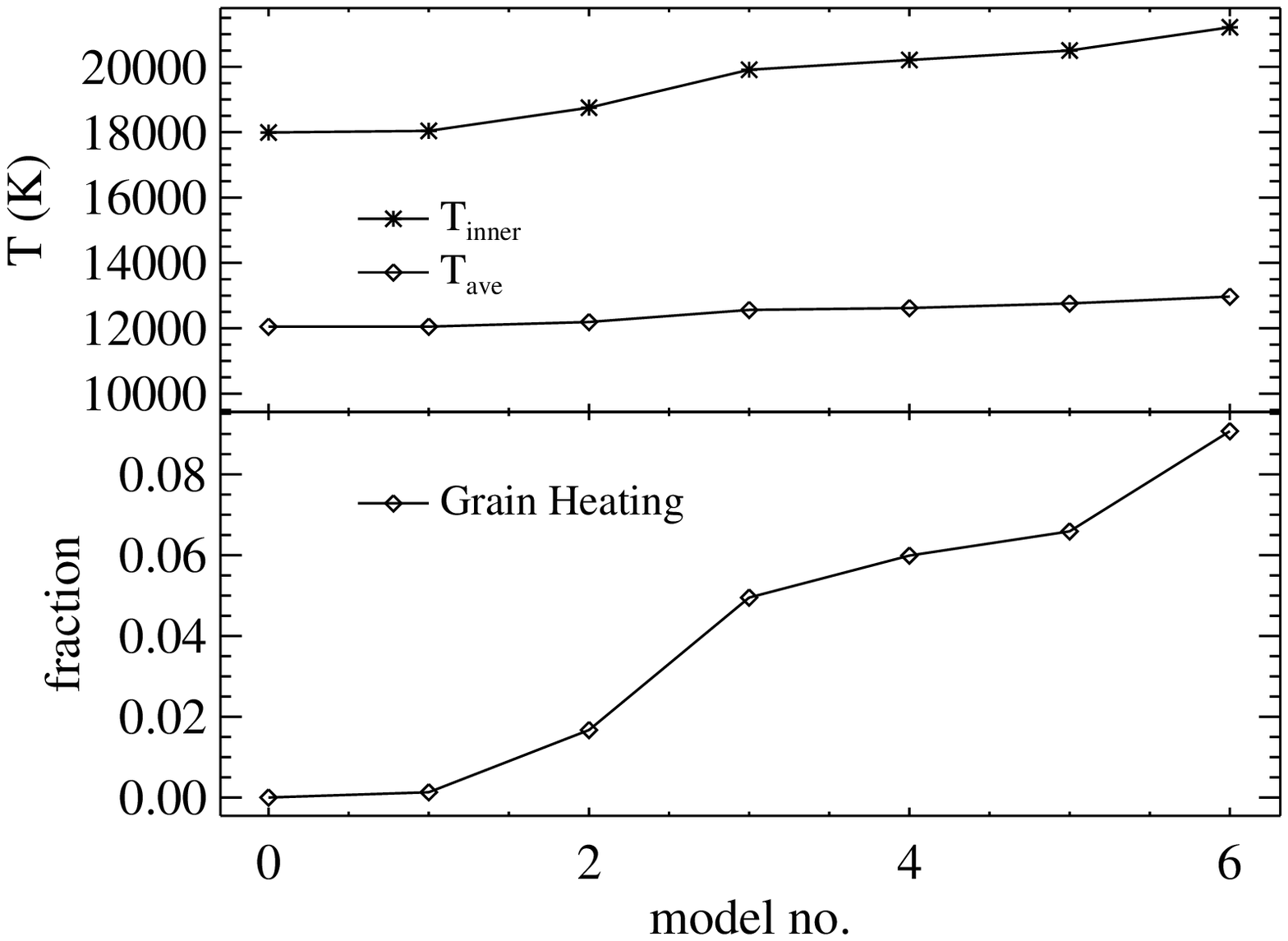}}
\caption{Same as Fig.~\ref{fig:hii}, but for the Paris PN models with silicate only.}
\label{fig:pn:sil}
\end{figure}

It is well known that photo-electric heating by grains in a photo-ionized
region has an important effect on the emitted spectrum (e.g., Volk 2001;
Dopita \& Sutherland 2000). It is however not well known that the size
distribution of the grains plays a very important role in determining the
magnitude of this effect. In order to test this, we have constructed a set of
models with Cloudy 96 beta~5 based on the standard Paris H\,{\sc ii} region
and planetary nebula (PN) models (P\'equignot, 1986). The Paris H\,{\sc ii}
region model would be roughly valid for a low-excitation PN as well, while the
Paris PN model is valid for a high-excitation PN. The base models contain no
dust, and will be used as a point of reference. We constructed 6 models out of
each base model by simply adding a dust component. We stress that in all six
models the chemical composition and the dust-to-gas mass ratio of the dust is
the same, and the only difference is the size distribution. Two models were
using single sized grains of 1.0 and 0.1~$\mu$m, while the other 4 were using
more or less realistic size distributions taken from the literature (as
indicated underneath the plots). We also point out that the models are
ionization bounded, so the outer radius varies, depending on the total opacity
of the grains (which also depends strongly on the size distribution). For the
Paris H\,{\sc ii} region model we added a mixture of graphite and silicates,
while for the Paris PN models we made separate models for graphite and
silicates since these grain types are not expected to co-exist. In
Figs.~\ref{fig:hii}, 2, and 3 we show the results of these calculation. In the
top-left panel we show the strength of selected infrared fine-structure lines.
These are expected to be mostly insensitive to electron temperature and
therefore show the difference in the overall ionization structure. In the
top-right panel we show optical/UV forbidden lines of the same species,
clearly showing that the enhancement for these lines is usually much stronger,
especially for highly excited lines which are only populated in the inner
regions of the nebula. This clearly shows the excess collisional excitation
caused by the photo-electric effect. In the bottom panel of each plot this is
further illustrated by showing the electron temperature at the inner edge, as
well as averaged over the ionized region, and the fraction of the total gas
heating that is due to the photo-electric effect.


All these plots clearly illustrate that the size distribution alone has a
dramatic effect on the emitted spectrum, and is therefore an important
parameter in the modeling of spectra from H\,{\sc ii} regions and PN.

\section{Conclusions}

In this paper we studied the effect that the distribution of grain sizes has
on the amount of photo-electric heating. We have shown that this effect is
nothing short of dramatic, making the grain size distribution an important
parameter in modeling of photo-ionized regions such as H\,{\sc ii} regions and
planetary nebulae. Only few studies of grain size distributions exist, and
they mainly concentrate on the diffuse interstellar medium (ISM) in order to
explain extinction curves. Further study of grain size distributions will be
needed in order to enable more accurate modeling of photo-ionized regions.
This is especially the case for planetary nebulae since it is not clear
whether ISM size distributions are valid for these objects.

\end{document}